\documentclass[conference]{IEEEtran}
\usepackage{cite}
\usepackage[cmex10]{amsmath}
\usepackage{graphicx}
\usepackage{subfigure}
\usepackage[update,prepend]{epstopdf}
\usepackage{amssymb,amsmath}
\usepackage{amsthm}
\begin{document}
\title{On the Impact of Relay-side Channel State Information on Opportunistic Relaying}

\author{
\IEEEauthorblockN{Anvar Tukmanov, Said Boussakta, Zhiguo Ding}
\IEEEauthorblockA{School of Electrical and Electronic Engineering,\\
Newcastle University, NE1 7RU, UK\\
Email: anvar.tukmanov@ncl.ac.uk}
\and
\IEEEauthorblockN{Abbas Jamalipour}
\IEEEauthorblockA{School of Electrical and Information Engineering\\
University of Sydney, NSW 2006, Australia\\
Email: a.jamalipour@ieee.org}}
\maketitle

\begin{abstract}
In this paper, outage performance of network topology-aware distributed opportunistic relay selection strategies is studied with focus on the impact of different levels of channel state information (CSI) available at relays. Specifically, two scenarios with (a) exact instantaneous and (b) only statistical CSI are compared with explicit account for both small-scale Rayleigh fading and path loss due to random inter-node distances. Analytical results, matching closely to simulations, suggest that although similar diversity order can be achieved in both cases, the lack of precise CSI to support relay selection translates into significant increase in the power required to achieve the same level of QoS. In addition, when only statistical CSI is available, achieving the same diversity order is associated with a clear performance degradation at low SNR due to splitting of system resources between multiple relays.
\end{abstract}

%
\IEEEpeerreviewmaketitle

\section{Introduction}
\subsection{Motivation and related work}
Cooperative relaying has been recognized as a cost-effective alternative to achieve diversity/multiplexing gains, by forming a virtual multi-antenna system from multiple single-antenna nodes. However, this reduction in requirements for individual nodes comes at a price of additional coordination overhead for the communication system. Capable of reducing such overhead, opportunistic relaying has been proposed in \cite{ble07}, as a technique that instead of coordinated transmission by $k$ relays utilizes only one relay with the best connection to the destination node. Such approach removes the need for tight coordination, while delivering the same diversity order and enabling distributed relay selection implementation, for instance, using timer-based methods.

For the original opportunistic relaying in \cite{ble07}, each relay needs an accurate estimate of current state of the channel between itself and the destination. Due to random channel fluctuations and additional complexity, imposed by the requirement to deliver CSI to transmitters, active research efforts have been concentrated on the analysis and design of opportunistic relaying with less dependence on precise CSI feedback. For example, the impact of available CSI on outage performance has been studied in \cite{mu:zd08} for the cases when the source node has access to different levels of information about the source-relay-destination path. In \cite{cui09}, a two-hop decentralized opportunistic relaying strategy with only local incoming CSI at receivers and index-valued CSI feedback at transmitters has been proposed and analyzed. Authors show that for a network of $n$ source-destination pairs and $m$ half-duplex non-cooperating relays, throughput of $m/2$ bps/Hz can be delivered through multiuser diversity. The impact of outdated CSI has been recently studied for opportunistic DF relaying in \cite{li11}.

However, current solutions generally rely on modeling of the fluctuation in inter-node links by small-scale fading only, without explicit account for propagation path loss due to spatial separation of communicating nodes. One common approach to incorporate the effect of path loss into the system model is to link the variance of small-scale fading with static inter-node distance \cite[p.73]{hong01}. Another approach is to explicitly account for the random node placement in the network, i.e. to account for network topology. Analytical basis for such approach can be developed using stochastic geometry methods \cite{Sgeom}. In \cite{far12} decentralized relay selection schemes for AF opportunistic relaying have been considered for the scenario where relays have access local-only CSI. However, only the special case of the destination located in the far-field was considered in \cite{far12}, so that the distance from the destination to all cooperating nodes was assumed to be identical.

The aim of this paper is to assess the impact of the level of available CSI on outage performance of a cooperative system with account for spatial dimension. Specifically, we consider two cases where relays have access to (a) exact instantaneous local CSI, both for the source-relay and for relay-destination links; and (b) exact instantaneous CSI for incoming links and only statistical CSI for the relay-destination links, which can be translated into ordering of relays with respect to distance to the destination. Note that relays operate in a distributed fashion, with no information exchange between relays.
Different from \cite{mu:zd08,ble07}, we take into account inter-node distances and randomness in node locations, such that the choice of the best relay now depends on signal attenuation due to propagation over random distance. On the other hand, arbitrary respective locations of the destination and cooperating relays are considered, such that the scenario in \cite{far12} can be viewed as a special case.

\subsection{Contributions and organization}
Contributions of this paper are twofold. First, we compare the effects of different levels of CSI available to relays and highlight the cost of limited CSI on required power consumption to achieve certain QoS. Second, for the scenario with statistical CSI, we provide a tractable analytical framework based on stochastic geometry, that accurately captures performance of such random cooperative system.
This paper is organized as follows. Network model, transmission strategy and CSI assumptions are presented in Section~\ref{sec:sysmodel}. Distributed relay selection strategies based on these assumptions are analyzed in Section~\ref{sec:strategies}. Outage performance results are presented and discussed in Section~\ref{sec:results} and Section~\ref{sec:concl} concludes the paper.

\section{System model}\label{sec:sysmodel}
\subsection{Network model}
\begin{figure}[t]
\centering
\includegraphics[width=\columnwidth]{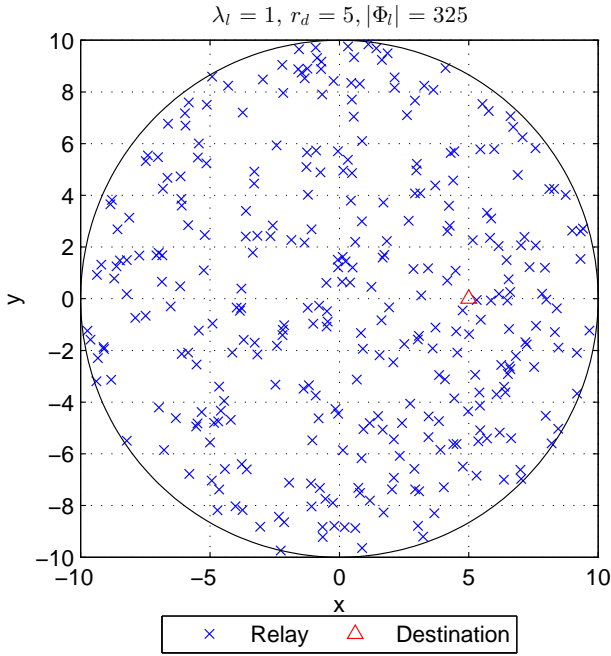}
\caption{Illustration of system setup: the BS at the center of the cell $W$ aims to communicate with a destination through one or $k$ randomly distributed relays.}
\label{fig:setup}
\vspace{-10pt}
\end{figure}

We consider a circular cell $W$ of radius $R$ with a base station (BS) acting as source $s$ at the cell center, one destination $d$ at distance $r_d$ from the BS and multiple randomly distributed relays constituting a realization of a homogeneous Poisson point process (PPP) $\Phi_l(W) = \left\{x_1,\ldots,x_l,\ldots,x_L\right\}$ with intensity $\lambda_l$. The number of relays $L = |\Phi_l(W)|$ in a realization of the point process is Poisson-distributed as 
\begin{equation}
\label{eq:ppp}
\text{Pr}\left(|\Phi_l(W)| = L\right) = e^{-\Lambda_l(W)} \frac{\left(\Lambda_l(W)\right)^{L}}{L!},
\end{equation}
where $\Lambda_l(W) = \int\limits_{W} \lambda_l(w) \text{d}w$ is the intensity measure of $\Phi_l(W)$, and $\lambda_l(w)$ is the intensity function of the process at some location $w$ in the cell $W$. Fig.~\ref{fig:setup} illustrates the setup.

\subsection{Signal model}\label{sec:tx}
Two time slot model is considered, as in \cite{ble07}. In the first time slot, the BS broadcasts a message, and all candidate relays in $\Phi_l(W)$ listen. Direct link between the source and the destination is assumed to be unavailable. The signal $u_l$ received by the  relay $x_l \in \Phi_l(W)$ can be written as
\begin{equation}
u_l = \sqrt{\frac{P_s}{1+r_{sl}^{\alpha}}} h_{sl}^{*} m + w_l,
\end{equation}
where $P_s$ is the source transmission power, $h_{sl}$ is the complex baseband Rayleigh fading coefficient, $r_{sl}$ is the distance between the source and the relay $x_l$, $\alpha$ is path loss exponent, $m$ is the transmitted message and $w_l$ is additive white Gaussian noise with variance $\sigma^2_{l}$.

The probability $p_l$ of successful reception of the source signal at the relay $x_l$ at a particular location $r_{sl}$ can be expressed as
\begin{equation}
\begin{aligned}
p_l &= \text{Pr}\left(\frac{|h_{sl}|^2}{1+r_{sl}^2} \geq \frac{2^{2\mathcal{R}}-1}{P_s/\sigma^{2}_{l}}\right)
		 =\mathbb{E}_{r_{sl}}\left\{e^{-\theta_l\left(1+r_{sl}^{\alpha}\right)}\right\}, 
\end{aligned}
\end{equation}
where $\mathcal{R}$ is required data rate, $\theta_l = \left(2^{2\mathcal{R}}-1\right)/\left(P_s/\sigma^{2}_{l}\right)$ is the threshold for correct decoding at the relay, and the expectation is taken with respect to the random distance $r_{sl}$ from the source to the relay. The relays that are able to decode the source message $m$ form a PPP $\Phi_j$ of relays qualified for retransmission:
\begin{equation}
\label{eq:phij}
\begin{aligned}
\Phi_j(W) = \left\{x_l \in \Phi_l(W) : \frac{|h_{sl}|^2}{1+r_{sl}^{\alpha}} \geq \theta_l \right\}. 
\end{aligned}
\end{equation}

Depending on available CSI at the relay and specific retransmission strategy for the second time slot, either one or $k$ relays will retransmit source message under the total power constraint of $P_s$ for the cooperative phase. When $k$ relays are scheduled for retransmission, equal power allocation between relays is considered, such that each relay retransmits with power $P_k = P_s/k$. Moreover, since the case of multiple retransmitting relays will require $k$ channel uses, each relay transmission must run at the rate of $\mathcal{R}_k = k \mathcal{R}$. Therefore, the threshold for correct decoding at the destination for the case of statistical CSI at relays is 
\begin{equation}
\begin{aligned}
\theta_k = \frac{k \left(2^{(1+k)\mathcal{R}}-1\right)}{P_s/\sigma^{2}_{l}},
\end{aligned}
\end{equation}
with the power $(k+1)$ is due to one channel use by the source and $k$ channel uses by the relays.

\subsection{CSI availability at relays}
We assume that all receiving nodes in the network can perfectly estimate the CSI for incoming signals, which are subject to small-scale Rayleigh fading and propagation path loss. The source node is assumed to have no information about any of $L$ source-relay links, whereas relays have access to CSI of either of the following two types:
\begin{enumerate}
	\item Exact instantaneous CSI for the relay-destination channel for each relay, i.e. $|h_{sl}|^2/\left(1+r_{sl}^2\right)$ is perfectly known at each qualified relay $x_j \in \Phi_j$;
	\item Statistics for CSI is available for relay-destination channel for each relay, i.e. $\mathbb{E}\left\{|h_{sl}|^2/\left(1+r_{sl}^2\right)\right\}$ is known at each qualified relay $x_j \in \Phi_j$.
\end{enumerate}

In the following we will study the performance of the second phase of relaying, where selected relay(s) will retransmit the source message to the destination based on available CSI.

\section{Distributed relay selection strategies}\label{sec:strategies}
When $\Phi_j(W) \neq \emptyset$, additional available information can be used to design relay selection strategy for the second time slot. The main focus of this section is on the case of statistical CSI, however we will first study the reference case with full CSI.

\subsection{Exact CSI based selection}\label{sec:full}
This reference scenario represents the case when perfect instantaneous CSI for individual relay-destination channels is made available to each relay. Clearly, communication outage in this case is only possible when there are no relays with reliable links both to the source and to the destination (this includes the case when there are no qualified relays, i.e. $\Phi_j(W) = \emptyset$). Therefore, it does not matter which of the relays with guaranteed connection to select from outage performance perspective (although one could find an energy-optimal solution).

Such relay qualification can be effectively modeled using thinning of point processes, as discussed in \cite{stoyan08,wan11TW}. Specifically, if a relay $x_j$ is located at distance $r_{j}$ from the source and at $r_{jd}$ from the destination respectively, then the probabilities of successful connection to the source and destination under the effects of small-scale Rayleigh fading and propagation path loss can be expressed respectively as
\begin{align}
p_{s} &= \mathbb{E}\left\{e^{-\theta_k}\left(1+r^{\alpha}_{j}\right)\right\}, \ 
p_{d} = \mathbb{E}\left\{e^{-\theta_k}\left(1+r^{\alpha}_{jd}\right)\right\}, \nonumber
\end{align}
where the expectations are taken over random distances $r_j$ and $r_{jd}$, and decoding thresholds at the relay and destination are same. Joint consideration of above criteria allows to approximate the mean number $\Lambda_q$ of relays that satisfy both connectivity conditions for cell radii $R\rightarrow \infty$ as
\begin{equation}
\label{eq:P1}
\begin{aligned}
\Lambda_q \approx \frac{\pi \lambda_l}{2\theta_k} e^{-\theta\left(2+r_d^2/2\right)},
\end{aligned}
\end{equation}
where $\lambda_l$ is the intensity function of the process of candidate relays; $\theta_k$ is the correct decoding threshold, assumed to be identical for the relay and the destination for the case of $k=1$; and $r_d$ is the distance from source to destination. The details of this derivation are omitted due to space limitation.

The number of points $Q$ in the process $\Phi_q(W)$ of relays connected both to the source and destination follows Poisson distribution. Therefore it is easy to see that the outage event for the case with full CSI at relays corresponds to the case when $Q=0$ and has probability given as
\begin{equation}
\begin{aligned}
P_{inst} &= \text{Pr}\left(\Phi_q(W) = \emptyset\right) = e^{-\Lambda_q(W)},
\end{aligned}
\end{equation}
where an approximation to $\Lambda_q(W)$ is given in \eqref{eq:P1}.

\subsection{Statistical CSI based selection}
This scenario corresponds to the case when the relays have access to local statistics of channels to the destination node, however no instantaneous CSI is available. Specifically, we consider the case where each relay knows $\mathbb{E}\left\{|h_{sl}|^2/\left(1+r_{sl}^2\right)\right\}$, and that based on these channel statistics, all relays are able to estimate the distance from the destination node with the \emph{same level of precision}. Note that no exact distance estimation is required, any technique sufficiently effective to correctly order relays with respect to the channel quality to the destination is acceptable (eg. \cite{far12}). The result of such distance-based ranking is an ordered sequence of relays $\{x_{(1)},\ldots,x_{(j)},\ldots,x_{(J)}\}$, where the relay $x_{(1)}$ has the largest distance to the destination, and relay $x_{(J)}$ -- the shortest. Such distributed ordering formation can be realized via timer-based algorithm, eg. \cite{far12}. 

Similarly to \cite{mu:zd08}, in the second time slot a set of $k$ relays with lowest distance estimates are selected from $\Phi_j(W)$ in a distributed fashion to forward the message to the destination. A number of assumptions will be made in the derivation of the outage probability $P_{stat}$ for this scenario:
\subsubsection{Outcome of source transmission} We assume that there are always enough qualified relays to meet the demand of $k$ relays, i.e. $|\Phi_j(W)| \geq k$. In practice the number of qualified relays $J = |\Phi_j(W)|$ may be less that the requested number of retransmitters $k$, or even be $0$. However for high SNR and small $k$, probability of such event can be shown to be small, while a precise account for $\Phi_j(W) < k$ would involve conditioning on a specific outcome of the PPP $\Phi_j(W)$, which in turn would require using analytically more complicated Binomial point processes for the parts of derivation \cite{stoyan08,ar1}. As will be verified by in Section~\ref{sec:results}, ignoring such technicality will lead to a minor mismatch between theory and simulations.
\subsubsection{Extension of $W$} Edge effects is a long-standing problem in spatial statistics \cite[p.132]{stoyan08}, associated with finite dimensions of the space where realizations of a point process take place. Consider our homogeneous PPP of candidate relays: strictly speaking, the points outside $W$ do not belong to $\Phi_l(W)$, which is why observations from the origin of $W$ will be different from those from the edge of $W$. While formally this is a contradiction to one of fundamental properties of a PPP, compensation for these effects increases analytical complexity. Fortunately for our scenario, impact of this formality is expected to be small because the number of points in $\Phi_j$ drops rapidly closer to cell edges (see \eqref{eq:phij}). Therefore, we will assume that the process $\Phi_j$ exists in the space beyond $W$, which will be shown to incur only a small mismatch between theory and simulation in Section~\ref{sec:results}.
 
Therefore, with the assumption of $|\Phi_j(W)| \geq k$, outage event $\mathcal{A}$ for this scenario can be expressed as 
\begin{equation}
\begin{aligned}
\mathcal{A} &= \bigcap\limits_{j=1}^{k}(\text{$j$-th nearest relay $x_{(j)}$ fails}).
\end{aligned}
\end{equation}
Note that the $k$ components of the set intersection above are mutually independent events, since both fading and placement of one node give no information about fading and placement of another (recall that $\Phi_j$ is a Poisson point process). Therefore, overall outage probability can be found as
\begin{equation}
\begin{aligned}
P_2 = \text{Pr}\left(\mathcal{A}\right) &=  \prod\limits_{j=1}^{k}{P_j},
\end{aligned}
\end{equation}
where $P_j$ is the probability that $j$-th nearest to the destination node relay will fail:
\begin{equation}
\label{eq:Pj}
\begin{aligned}
P_{j} &= \text{Pr}\left(\frac{|h_{jd}|^2}{1+r^{\alpha}_{jd}} < \theta_k \right) = 1- \mathbb{E}_{r_{jd}}\left\{e^{-\theta_k\left(1+r^{\alpha}_{jd}\right)}\right\} \\
&=1-\int\limits_{0}^{R+r_x} e^{-\theta_k\left(1+r^{\alpha}_{jd}\right)} f_{k} \left(r_{jd}\right) \text{d}r_{jd},
\end{aligned}
\end{equation}
where $h_{jd}$ is the complex baseband Rayleigh channel coefficient, $r_{jd}$ is the distance from the BS to the nearest relay $x_j \in \Phi_j$, and $f_{k} \left(r_{jd}\right)$ is the probability density function (PDF) for the distance to the $k$-th nearest to the destination relay. Using properties of Poisson point processes , the PDF $f_{k} \left(r_{jd}\right)$ can be given as \cite{stoyan08}
\begin{equation}
\label{eq:frcd}
\begin{aligned}
f_{k} \left(r_{jd}\right) = e^{-\Lambda_j^{'}\left(B\right)} \cdot \frac{2 \left(\Lambda_j^{'}\left(B\right)\right)^{k} }{r_{jd} \Gamma (k)}.
\end{aligned}
\end{equation}
Here $\Lambda_j^{'}\left(B\right)$ denotes the mean number of points of the PPP $\Phi_j$ of relays that can decode the source message inside a region $B \subseteq W$ with radius $r_{jd} \in [0,R+r_d]$, centered at the destination location. Our second assumption is used here, as formally $B$ cannot have circular shape, as $\Phi_j(W)$ does not span beyond $W$.

Both $\Lambda_j^{'}\left(\cdot\right)$ and $\Lambda_j\left(\cdot\right)$ are mean measures of the same Poisson point process $\Phi_j$ with location-dependent intensity function $\lambda_j(w)$, with the key difference in the position of observation points. Specifically, for $\Lambda_j\left(\cdot\right)$, the observation point is located at the BS, so that while the resulting PPP $\Phi_j$ is inhomogeneous, it is still isotropic with respect to the BS. On the other hand, $\Lambda_j^{'}\left(\cdot\right)$ measures the number of points of the same process but from the destination point of view, which makes PPP $\Phi_j$ anisotropic from such perspective. Indeed, when observed from the destination, it is more likely to find relays with reliable connections to the BS at angles $\varphi_{jd}$ pointing towards the BS, rather than in the opposite direction.

The mean number $\Lambda_j^{'}\left(B\right)$ of relays connected to the BS falling within a circular region $B$ with radius $r_{jd}$ can be expressed in terms of location-dependent, but universal intensity function $\lambda_{j}(w)$ as
\begin{equation}
\label{eq:Lj}
\begin{aligned}
\Lambda_j^{'}\left(r_{jd}\right) = \int\limits_{B} \lambda_{j}(w)\text{d}w = \int\limits_{0}^{2\pi} \int\limits_{0}^{r_{jd}} \lambda_{j}(r_j)r\text{d}r\text{d}\varphi,
\end{aligned}
\end{equation}
where $r_j$ is the distance from the BS to a relay $x_j$. The intensity function $\lambda_{j}(w)$ of the process of relays connected to the BS can be expressed as \cite{stoyan08}
\begin{equation}
\begin{aligned}
\lambda_{j}(w) = \lambda_l p_j(w) = \lambda_l e^{-\theta(1+r_j^{\alpha})}.
\end{aligned}
\end{equation}

Using standard trigonometry, we can rewrite $\lambda_{j}(r_j)$ in terms of integration variables $\lambda_{j}(r,\varphi)$
\begin{equation}
\begin{aligned}
\lambda_{j}(r,\varphi) = \lambda_l e^{-\theta\left(1+r_d^2 + r^2 -2 r_d r \cos(\varphi)\right)}, 
\end{aligned}
\end{equation}
which after substitution into \eqref{eq:Lj} gives 
\begin{equation}
\label{eq:Lj2}
\begin{aligned}
\Lambda_j^{'}\left(r_{jd}\right) = \lambda_l e^{-\theta\left(1+r_d^2\right)} \int\limits_{0}^{2\pi} \underbrace{\int\limits_{0}^{r_{jd}} r e^{-\theta\left(r^2 -2 r_d r \cos(\varphi)\right)} \text{d}r }_{I}\text{d}\varphi.
\end{aligned}
\end{equation}
Unlike estimation of outage probability in Section~\ref{sec:full}, approximation for large $r_{jd}$ is inapplicable, because we are interested in the behavior of $\Lambda_j^{'}\left(B\right)$, including for small $r_{jd}$ values. For this reason, we first take the inner integral to allow for subsequent numerical evaluation of \eqref{eq:Lj2} with respect to rotation angle $\varphi$ around the destination node. After straightforward, but lengthy integration, $I$ can be written as
\begin{equation}
\label{eq:I}
\begin{aligned}
I &= \frac{1}{2\theta_k}\left(1-e^{-\theta_k\left(r_{jd}^2 - a r_{jd}\right)}\right) \\
	&+ \sqrt{\frac{\pi}{\theta_k}} \frac{a}{4} e^{\frac{a^2}{4} \theta_k} \left(\text{erf}\left(\frac{a}{2} \sqrt{\theta_k}\right) - \text{erf}\left(\frac{a}{2} \sqrt{\theta_k}-\sqrt{\theta_k} r_{jd}\right)\right),
\end{aligned}
\end{equation}
where $a = 2 r_d \cos(\varphi)$ and $\text{erf}(\cdot)$ denotes error function. It is interesting to note that when the displacement of the observation point $r_d=0$, i.e. when the locations of the remote observation point and the BS coincide, \eqref{eq:I} reduces to $I = \left(1-e^{-\theta r_{jd}^2}\right)/2\theta_k$. However for general $r_{jd} \in [0,R+r_d]$, closed form solution of \eqref{eq:Lj2} can be overly complicated, and numerical solution will be used to obtain outage performance.
%

\section{Results and discussion}\label{sec:results}
First, we highlight the difference in spatial distribution of the relays connected to the BS, when observed from the BS itself and from the destination at distance $r_d$. The aims of such comparison are to (a) illustrate the impact of inhomogeneous structure of the process $\Phi_j(W)$ on the number of relays that can participate in cooperative transmission from different viewpoints and (b) verify the result in \eqref{eq:Lj}. Specifically, we are interested in the mean number of relays in the process $\Phi_j(W)$ that fall inside a circular region with radius $r$ centered either at the BS, or at the destination location. 

\begin{figure}[t]
\centering
\includegraphics[width=\columnwidth]{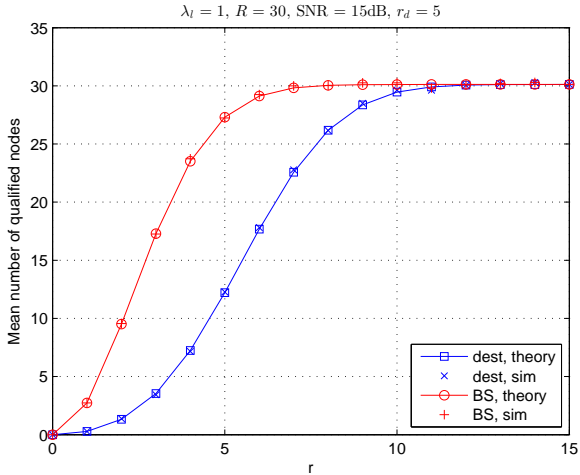}
\caption{Mean number of relays connected to the BS as a function of radius $r$, observed from the BS (red circles) and the destination at $r_d$ (blue squares).}
\label{fig:Intensity}
\vspace{-14pt}
\end{figure}

Fig.~\ref{fig:Intensity} illustrates the quantity of interest as a function of radius $r$  for transmit SNR of 15 dB and destination node located at $r_d = 5$. Note that the destination can find a smaller number of qualified relays within the same proximity compared to the BS, which is due to exponential decay in the received power as the source-relay distance increases, eg. the destination can expect support from $1$ qualified relay within $r=2$ on average. As $r$ grows, the circular region around the destination will eventually include all qualified relays, which can be seen from convergence of the curves for larger $r$. 
 
\begin{figure}[t]
\centering
\includegraphics[width=\columnwidth]{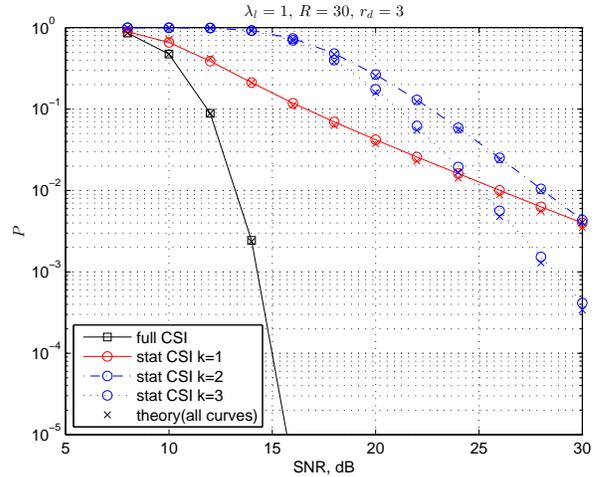}
\caption{Outage probability for opportunistic relaying from the BS to the destination for the cases of exact and statistical CSI at relays.}
\label{fig:Pout}
\vspace{-14pt}
\end{figure}

Finally, we study the impact of the level of available CSI on outage performance of opportunistic relaying. Fig.~\ref{fig:Pout} illustrates outage probability for the communication between the BS and the destination through either one best relay or $k$ nearest to the destination qualified relays from $\Phi_j$. As can be seen from the graph, outage probability drops as transmit SNR grows for the case of full CSI, which follows the intuition that with larger transmit power it becomes less likely that none of the relays will be connected both to the source and the destination. Correspondingly, the rate of such decay increases with growth in the expected number of qualified relays $\Lambda_q$. 
However, when the relays cannot decide whether their transmission will succeed or not due to the absence of instantaneous CSI, outage performance degrades significantly. In particular, when a single closest qualified relay retransmits the source message, the diversity order of the studied system becomes one irrespective of available power, as can be seen from Fig.~\ref{fig:Pout}. Such increased power requirement to meet certain QoS level can be viewed as penalty for lack of information. 

In order to improve performance the BS may ask $k>1$ nearest qualified relays to retransmit the message. Performance of such scenarios is illustrated by dashed lines in Fig.~\ref{fig:Pout}. Clearly, for sufficiently high SNR levels multiple retransmitting relays outperform single transmitting relay; on the other hand, for lower SNRs splitting power between relays leads to performance degradation. Indeed, larger $k$ values mean that each relay can be allocated less power from the budget. In addition, since relay retransmissions take $k$ additional time slots instead of $1$, outage performance is further degraded by $k$-fold increase in the required data rate for each relay transmission. Interestingly, the gain from increased diversity overweights the effect of resource splitting between $k$ relays relatively quickly for $k=3$, delivering better outage performance for lower power consumption as compared to $k=2$.

In summary, when instantaneous CSI is unavailable, increasing the number of active relays can lead to identical outage behavior as for the case of perfect CSI at the cost of higher consumed power. Therefore, it may be beneficial to employ more relays when power budget is sufficiently high, rather than to invest all power into one nearest relay transmission when instantaneous CSI is unavailable.

\section{Conclusion}\label{sec:concl}
In this paper we have studied the impact of available CSI at relays on outage performance of distributed opportunistic relaying. For the cases when either exact instantaneous or statistical CSI is available, outage probability performance has been compared with an explicit account for spatial placement of all participating nodes. Provided analytical and matching simulation results show that although identical diversity order can be achieved in both cases, delivering the same level of QoS in terms of outage probability for the case of statistical CSI at relays requires significantly larger power. Moreover, when only statistical CSI is available, increasing the number of retransmitting relays creates a trade-off between sacrificing performance at low SNR due to splitting resources among $k$ relays, and larger diversity order for high SNR.

%
%

\end{document}